\documentclass[final,1p,times]{elsarticle}

\usepackage{graphicx}

\usepackage{amssymb}
\usepackage{color}
\usepackage{psfrag,rotating}

\newcommand{\be}{\begin{equation}}
\newcommand{\ee}{\end{equation}}
\newcommand{\bea}{\begin{eqnarray}}
\newcommand{\eea}{\end{eqnarray}}

\journal{New Astronomy}

\begin{document}

\begin{frontmatter}



\title{Fermionic warm dark matter produces galaxy cores in the observed scales
because of quantum mechanics}


\author[1]{C. Destri}

\author[2,3]{H. J. de Vega\fnref{xx}}

\author[3]{N. G. Sanchez}

\address[1]{Dipartimento di Fisica G. Occhialini, Universit\`a
Milano-Bicocca and INFN, sezione di Milano-Bicocca, Piazza della Scienza 3,
20126 Milano, Italia.}

\address[2]{LPTHE, Universit\'e
Pierre et Marie Curie (Paris VI),
Laboratoire Associ\'e au CNRS UMR 7589, Tour 13-14, 4\`eme. et 5\`eme. \'etage, 
Bo\^{\i}te 126, 4, Place Jussieu, 75252 Paris, Cedex 05, France}

\address[3]{Observatoire de Paris, LERMA, Laboratoire
Associ\'e au CNRS UMR 8112,
61, Avenue de l'Observatoire, 75014 Paris, France}

\begin{abstract}
We derive the main physical galaxy properties: mass, halo radius, phase space density and velocity dispersion
from a semiclassical gravitational approach in which fermionic 
WDM is treated quantum mechanically. They turn out to be fully compatible  with observations.
The Pauli Principle implies for the fermionic DM phase-space density 
$ Q({\vec r}) = \rho({\vec r})/\sigma^3({\vec r}) $ the quantum bound $ Q({\vec r})  \leq 
   K \; m^4/\hbar^3 $, where $  m $ is the DM particle mass, $ \sigma({\vec r}) $ is the DM 
velocity dispersion and $ K $ is a pure number of order one which we estimate. 
Cusped profiles from $N$-body galaxy simulations produce a divergent $ Q(r) $ at $ r = 0 $ 
violating this quantum bound. The combination of this quantum bound with 
the behaviour of $ Q(r) $ from simulations, the virial theorem  
and galaxy observational data on $ Q $ implies lower bounds on the halo radius and a minimal 
distance $ r_{min} $ from the centre  at which classical galaxy dynamics for DM fermions breaks down. 
For WDM, $ r_{min} $ turns to be in the parsec scale.
For cold dark matter (CDM), $ r_{min} $ is between dozens of kilometers 
and a few meters, astronomically compatible with zero.
For hot dark matter (HDM), $ r_{min} $ is from the kpc to the Mpc. 
In summary, this quantum bound rules out the presence of 
galaxy cusps for fermionic WDM, in agreement with astronomical observations, which show that the DM
halos are cored. 
We show that compact dwarf galaxies are natural quantum macroscopic objects supported against 
gravity by the fermionic WDM quantum pressure (quantum degenerate fermions) 
with a minimal galaxy mass and minimal velocity dispersion. Quantum mechanical calculations 
which fulfil the Pauli principle become necessary to compute galaxy structures at kpc scales and below. 
Classical $N$-body simulations are not valid at scales below  $ r_{min} $.
We apply the Thomas-Fermi semiclassical approach to fermionic WDM galaxies, we resolve it numerically 
and find the physical galaxy magnitudes: mass, halo radius, phase-space density, 
velocity dispersion, fully consistent with observations especially for compact dwarf galaxies.
Namely, fermionic WDM treated quantum mechanically, as it must be, reproduces
the observed galaxy DM cores and their sizes. The lightest known dwarf galaxy (Willman I) 
implies a lower bound for the WDM particle mass $ m > 0.96 $ keV. 
These results and the observed galaxies with halo radius $ \geq 30 $ pc and halo mass  
$ \geq 4 \times 10^5 \; M_\odot $ provide further indication that the WDM particle mass
$ m $ is approximately in the range 1-2 keV.
\end{abstract}


\fntext[xx]{Corresponding author. H. J. de Vega, devega@lpthe.jussieu.fr, Tel. 33-1-4427-7394, Fax 33-1-4427-7393}

\end{frontmatter}


\section{Introduction and summary of results}

Dark matter (DM) is the main component of galaxies, especially of dwarf
galaxies which are almost exclusively formed by DM. It thus appears that studying
galaxy properties is an excellent way to disentangle the nature of DM.

\medskip

Cold DM (CDM) produces an overabundance of substructures below the $ \sim 50 $ kpc till very small scales  
$ \sim 0.005 $ pc which constitutes, as is well known, one of the most serious drawbacks for CDM.
On the contrary, warm DM (WDM),
that is, DM particles with mass in the keV scale, produces DM structures in the observed range
of scales $ \gtrsim 50 $ kpc. In WDM  structure formation,  substructures below the free-streaming
scale $ \sim 50 $ kpc are not formed contrary to the case of CDM.
This conclusion for WDM based on the linear theory is confirmed by
$N$-body simulations \cite{simuw1,simuw2,simuw3,simuw4,simuw5,simuw6,simuw7}. 
For scales larger than $ ~ 50 $ kpc, WDM yields the same results than CDM and agrees with all the observations.

\medskip

Astronomical observations show that the DM galaxy density profiles are {\bf cored} 
till scales below the kpc \cite{obs1,obs2,obs3,obs4,gil,wp}.
On the other hand, $N$-body CDM simulations exhibit cusped density profiles with a typical $ 1/r $ behaviour
near the galaxy center $ r = 0 $. WDM simulations exhibit cusps or small
cores smaller than the observed cores \cite{coreswdm1,coreswdm2,coreswdm3,coreswdm4}.

\medskip

Numerical calculations on the spherically symmetric Vlasov--Poisson
     equation based on the Larson moment expansion \cite{larson1,larson2}, as well as on the exact
     dynamics of the associated N-body system, have confirmed these
     findings \cite{nos2}.

\medskip

For fermionic DM the Pauli principle states that the phase-space distribution function
 for spin-$\frac12 $ particles $ f({\vec r},{\vec p}) $ must be smaller than two
\be\label{csup}
 f({\vec r},{\vec p}) \leq 2 \; .
\ee
Since the matter density $ \rho({\vec r}) $ is obtained from the phase-space distribution
through
\be
\rho({\vec r}) = m \; \int d^3p \; \frac{f({\vec r},{\vec p})}{(2 \, \pi \; \hbar)^3}
\quad ,  \quad m = {\rm DM ~ particle ~ mass} \; ,
\ee
this implies a bound on the phase-space density $ Q({\vec r}) \equiv \rho({\vec r})/\sigma^3({\vec r}) $
where $ \sigma({\vec r}) $ is the DM velocity dispersion, which
 takes the form (see sec. \ref{cotaq})
\be\label{cotI}
\frac{\hbar^3}{m^4} \;  Q({\vec r}) \leq K  \; .
\ee
Here $ K $ is a pure number of order one that we estimate in appendix \ref{calK} and display
in Table \ref{valK}. 
In the classical physics limit $ \hbar \to 0 $, the right hand side
goes to infinity and the bound on $ Q({\vec r}) $ disappears. 

\medskip  

The quantum bound eq.(\ref{csup}) in the
cosmological context has been  previously considered in refs. \cite{st,gorub,gorubL}
to derive lower bounds on neutral lepton masses. 
In the present paper, we use this bound in a new way 
incorporating the behaviour of the phase-space density  $ Q(r) $  \cite{espf1,espf2,espf3,espf4,espf5}
and observational data to derive lower bounds on the halo radius. These bounds are of
semiclassical gravitational nature and motivate us to use the Thomas-Fermi approach for galaxies,
namely, treating gravitation classically and fermionic WDM quantum mechanically.

We combine the quantum bound eq.(\ref{cotI}) with the results from
classical $N$-body simulations which indicate a simple power law behaviour for $ Q(r) $ 
\cite{espf1,espf2,espf3,espf4,espf5}
\be\label{qsI}
 Q(r) = Q_h \; \left(\frac{r}{r_h}\right)^{-\beta}
\ee
where $ \beta \simeq 1.9 - 2 , \; r_h $ is the halo radius and $ Q_h $
stands for the mean phase space density in the halo.
Classical WDM $N$-body simulations yield results similar to the behaviour
eq.(\ref{qsI}) even for $ r \lesssim  r_h $, when properly using the velocity dispersion of the
keV-scale WDM particles \cite{sinziana,shao}. 

Classically, $ Q(r) $ grows unbounded for $ r \to 0 $ 
violating the quantum bound eq.(\ref{cotI}).
Therefore, for small enough scales classical physics in galaxies {\bf breaks down}. 
In other words, classical $N$-body simulations and hydrodynamical simulations
are not applicable near the galaxy centers.

\medskip

We show that the classical result eq.(\ref{qsI}) only holds for distances
far enough from the galaxy center, namely for
\be\label{rqI}
r \geq r_{min} \equiv \frac{\hbar^{\frac32}}{m^2} \; \sqrt{\frac{Q_h}{K}} \; r_h \; .
\ee
For a DM particle mass in the keV scale and the observed values of $ r_h $ and $ Q_h $,
it turns out that $ r_{min}  $ can be between $ 0.1 $ and $ 1 $ pc.

\medskip

Eq.(\ref{rqI}) provides the {\bf minimal} distance from the center where quantum mechanical
effects are {\bf essential} and rules out the presence of galaxy cusps for fermionic WDM.
Moreover,  quantum mechanical effects extend well beyond $ r = r_{min}  $ 
because smoothing up the classical cusp which sucks matter 
towards the origin has an effect on the whole galaxy halo.

Taking into account the quantum nature of fermionic DM introduces
a physical length scale of quantum origin. As a consequence,  density profiles
become regular (cored) at the origin. For bosonic DM, the bound eq.(\ref{cotQ}) does not apply
and the formation of cusps is allowed. However, the observed galaxy density profiles 
are cored \cite{obs1,obs2,obs3,obs4,gil,wp}.

\bigskip

The quantum bound eq.(\ref{cotQ}) applies for any kind of  fermionic dark matter with the value of
$ r_{min} $ determined by the DM particle mass. For CDM where 1 GeV $ < m < $ 100 GeV,
$ r_{min} $ is between dozens of kilometers and a few meters, astronomically compatible with zero.
Therefore, classical approaches as $N$-body simulations and the classical
Boltzmann-Vlasov equation fully apply for CDM and unavoidably produce cusps.

\medskip

HDM where 1 eV $ < m < $ 10 eV, suppresses all structures scales below the Mpc scale due to its long
free-streaming length and has been ruled out years ago.
Anyway, we find that classical aproaches to fermionic HDM as
$N$-body simulations are not valid for scales below an $ r_{min} $ which
turns to be between the kpc and the Mpc, depending on the galaxy type.

\medskip

A direct way to see whether a system of particles has a classical or quantum nature
is to compare the particle de Broglie wavelength $ \lambda_{dB} $ with the interparticle distance $ d $.
We do that in sec. \ref{broglie} and express their ratio as 
$$
{\cal R} \equiv \frac{\lambda_{dB}}{d} = \hbar \; \left( \frac{Q_h}{m^4}\right)^\frac13 \; .
$$
The observed  values of $ Q_h $ from Table \ref{pgal} yields
\be
2 \times 10^{-3}  < {\cal R} \; \left( \frac{m}{\rm keV}\right)^\frac43 < 1.4 \; .
\ee
The larger value of $ \cal R $ is for ultracompact dwarfs and the smaller value of $ \cal R $ 
is for big spirals. The values of $ \cal R $ around unity clearly imply
(and solely from observations) 
that compact dwarf galaxies are natural {\it macroscopic quantum objects} for WDM.

\medskip

WDM fermions always provide at least a pressure of quantum nature. When
this quantum pressure is balanced with the gravitation pressure, we find 
values for the total mass $ M \sim 10^6 \;  M_\odot$, the radius $ R \sim 30 $ pc and the velocity dispersion
$ \sigma \sim 2 $ km/s consistent with compact dwarf galaxies (see Table \ref{pgal}). 
These results back the idea that dwarf spheroidal galaxies are supported by the
fermionic {\it WDM quantum pressure}.

\medskip

We then treat the self-gravitating fermionic DM in the Thomas-Fermi approximation.
In this approach, the DM chemical potential $ \mu(r) =  \mu_0 - m \; \phi(r) $,
where $ \mu_0 $ is a constant and $ \phi(r) $ the gravitational potential,
obeys the Poisson equation
\be\label{poisI}
\frac{d^2 \mu}{dr^2} + \frac2{r} \; \frac{d \mu}{dr} = - 
\frac{4 \, \pi \; G \; m^2}{\pi^2 \; \hbar^3} \int_0^{\infty} p^2 \; dp \; f\left(e(p)-\mu(r)\right)
\ee
where $ G $ is Newton's gravitational constant, $ p $ is the DM particle momentum,
$ e(p) = p^2/(2 \, m) $ is the DM particle kinetic energy and $ f(E) $ is the
energy distribution function, 

This is a semiclassical gravitational approach to determine selfconsistently the
gravitational potential of the fermionic WDM given its distribution function $ f $.

\medskip

In order to have  bounded DM mass densities according to eq.(\ref{cotI})
we impose the boundary condition at the origin: $ \mu'(0) = 0 $.

The distribution function $ f(E) $ is determined by the DM evolution
since decoupling. Such evolution must take into account the quantum character
that emerges when the distribution function approaches the quantum upper bound eq.(\ref{cotI}).
Such quantum dynamical calculation is beyond the scope of the present paper.
We modelize here the distribution function $ f(E) $ by the equilibrium
Fermi-Dirac distributions and by out of equilibrium distributions for sterile
neutrinos  (see the appendix \ref{calK}).

We get a one parameter family of solutions of eqs.(\ref{poisI})
parametrized by the value of the chemical potential at the origin  $ \mu(0) $.
We then express the chemical potential at the origin in terms of the phase-space density
at the origin $ Q(0) $. 
Large positive values of 
$ \mu(0) $ correspond to the quantum degenerate 
fermions limit, while large negative values of $ \mu(0) $ 
yield the dilute (classical) limit. 

We show that the Thomas-Fermi equation implies the local equation of state
$$
P(r) =  \frac13 \; v^2(r) \; \rho(r) \;  \; ,
$$
and the hydrostatic equilibrium equation
$$
\frac{dP}{dr} + \rho(r) \; \frac{d\phi}{dr} = 0 \; .
$$
This local equation of state generalizes the local perfect fluid equation 
for $r$-dependent velocity $ v(r) $.

\medskip

In the quantum degenerate fermions limit, the halo radius, the velocity dispersion and the galaxy mass 
take their {\it minimum} values. 
These minimum values are similar to the estimates for degenerate fermions given in sec. \ref{qup}. 

The masses of compact dwarf spheroidal galaxies dominated by DM must be 
larger than this minimum mass $ M_{min} $.
The lightest known  galaxy of this kind is Willman I (see Table \ref{pgal}).
Imposing $ M_{min} < M_{Willman ~ I} =  3.9 \; 10^5 \; M_\odot $ provides a lower bound for the WDM particle mass:
\be
m > 0.96 \; {\rm keV} \; . 
\ee
The numerical resolution of eqs.(\ref{poisI}) for the whole range of
the chemical potential at the origin  $ \mu(0) $ yields the physical
galaxy magnitudes: mass, halo radius, phase-space density and velocity dispersion fully compatible
with observations especially for compact dwarf galaxies as can be
seen from figs. \ref{RQhalo} and \ref{Mhalo} and Table \ref{pgal}.

Approaching the classical diluted limit yields larger and larger halo radii, galaxy masses
and velocity dispersions. Their maximum values are limited by the initial conditions
provided by the primordial power spectrum which determines the sizes and masses of the galaxies formed.

The phase space density decreases from its maximum value for the compact dwarf galaxies 
corresponding to the degenerate fermions limit till its smallest value for large galaxies
(spirals and ellipticals) corresponding to the classical dilute regime.

The theoretical values for $ r_h , \; M $ and $ v(0) $ from the Thomas-Fermi approach 
vary very little with the specific form of the phase-space distribution function 
$ f(E) $  as function of the energy (Fermi-Dirac distribution
or out of equilibrium sterile neutrino distribution).

Comparison of the theoretically derived galaxy masses with the galaxy data in fig. \ref{Mhalo} indicates 
a WDM particle mass $ m $ approximately in the range 1 - 2 keV. For larger masses, $ m \gg 1 $ keV 
an overabundance of small galaxies (small scale structures) without observable counterpart appears.

\medskip

In summary, the theoretical Thomas-Fermi results are fully consistent with all the observations especially
for dwarf compact galaxies as can be seen from figs. \ref{RQhalo} and \ref{Mhalo}.
It is highly remarkably that in the context of fermionic WDM the simple static
quantum description provided by Thomas-Fermi is able to reproduce such broad variety of galaxies.

\medskip

These results indicate that fermionic WDM treated quantum mechanically (even approximately)
is fully consistent with the observed galaxy properties including the DM core sizes.

Therefore, the effect of including baryons is expected to be a correction to the pure WDM results
presented in this paper, consistent with the fact that dark matter is in average six times more abundant than baryons.

\medskip

In section \ref{cotaq} we describe the quantum bound for the phase-space density of fermionic DM
and we derive the relevant galaxy scales where DM classical physics in galaxies breaks down. 
In section III we show how the quantum fermionic WDM pressure balances the gravitational pressure in galaxies. 
In section IV
the Thomas-Fermi approach for self-gravitating fermionic WDM is presented
and applied to galaxies, showing that fermionic WDM treated quantum mechanically is able to reproduce
the observed DM properties of galaxies including the DM cores and their sizes. 
We derive in Appendix A the numerical values of the constant
$ K $ in the quantum bound for several momentum distributions.

We use units where the speed of light is taken $ c = 1 $ along this paper.

\section{Quantum bounds for fermionic DM in galaxies from the Pauli 
principle}\label{cotaq}

For fermionic DM the Pauli principle tells us that the 
number of spin states in a phase-space cell volume $ (2 \, \pi \; \hbar)^3 $
cannot be larger than $ 2 $ for spin-$\frac12 $ particles. Namely, the phase space distribution function
$ f({\vec r},{\vec p}) $ must satisfy
\be\label{pauli}
 f({\vec r},{\vec p}) \leq 2 \; .
\ee

The DM number density can be expressed as
\be\label{n}
n({\vec r}) = \int d^3p \; \frac{f({\vec r},{\vec p})}{(2 \, \pi \; \hbar)^3}
=  \frac{m^3}{2 \; \hbar^3} \; \sigma^3({\vec r}) \; {\bar f}({\vec r}) \; K \; ,
\ee
where $ {\bar f}({\vec r}) $  is the maximal ${\vec p}-$average of the
phase space distribution over a volume $ m^3 \; \sigma^3({\vec r})$,
$ m $ is the mass of the DM particle, $ \sigma({\vec r}) $ is the DM velocity dispersion,
$ \sigma^2({\vec r}) \equiv <v^2({\vec r})> / 3 $ 
and $ K $ is a pure number of order one.
In appendix \ref{calK} we evaluate $ K $ for distributions of cosmological
relevance and display it in Table \ref{valK}.

We find from eq.(\ref{n})
\be\label{f}
{\bar f}({\vec r}) = \frac{2 \;  \hbar^3}{K \; m^3} \; \frac{n({\vec r})}{\sigma^3({\vec r})}
\ee
Inserting eq.(\ref{f}) into eq.(\ref{pauli}) which also applies to $ {\bar f}({\vec r}) $ yields,
\be\label{cota}
\frac{n({\vec r})}{\sigma^3({\vec r})} \leq  \frac{K \; m^3}{\hbar^3}\; .
\ee
As a consequence, we find the following bound for the phase-space density:
\be\label{cotQ}
Q({\vec r}) \equiv \frac{\rho({\vec r})}{\sigma^3({\vec r})} \leq  K \; 
\frac{m^4}{\hbar^3} \; .
\ee
where $ \rho({\vec r}) = m \; n({\vec r}) $ is the matter density.
Therefore, the phase space density $ Q({\vec r}) $ can {\bf never} take values
larger than the right hand side of eq.(\ref{cotQ}).
This is an absolute quantum upper bound which is due to quantum physics, namely
the Pauli principle. 

\begin{table}
\begin{tabular}{|c|c|c|c|} \hline  
 & &  & \\
Distribution Function  & $ \frac1{\displaystyle \psi(0)} $ & $ K $ & 
$ \frac{\displaystyle 0.60364}{\displaystyle \sqrt{K}} $ \\ & & & 
\\ \hline  &  &  &
   \\  Fermi-Dirac and DW & $ 3 \; \zeta(3) = 3.60617 $ & $ 1.89858 $ & $ 0.43809 $ \\  
 &  &  & \\ \hline  &  &  &
 \\ $\nu$-MSM & $ 0.045 \; \pi \;\zeta(5) = 0.14659 $ & $ 0.077178 $ & $ 2.17285 $ \\  
&  &  & \\ \hline  &  &  &
\\ Maxwell-Boltzmann & $ \sqrt{\displaystyle \frac{\pi}2} = 1.2533 $ & $ 0.65985 $ & $ 0.74311 $ \\ 
 & & & \\ \hline  
\end{tabular}
\caption{The constant $ K $ defined by eq.(\ref{cotQ}) and $ 0.60364/\sqrt{K} $.
We see that the characteristic size is larger for sterile neutrinos decoupling
out of equilibrium.} 
\label{valK}
\end{table}

\medskip

In the classical physics limit $ \hbar \to 0 $ the right hand side of eq.(\ref{cotQ}) 
tends to infinity and the bound disappears. 

\medskip

Since the squared velocity dispersion cannot be larger than the speed of light
we have a lower bound for the phase-space density
$$
Q \geq Q_{min} \equiv 3 \; \sqrt3 \; \rho \; .
$$

The upper bound  eq.(\ref{cotQ}) on $ Q $ implies, at given density $ \rho $, a lower 
bound $ v_{min} $ on the velocity 
$$
Q_{max} = 3 \; \sqrt3 \; \frac{\rho}{v_{min}^3} =  K \; 
\frac{m^4}{\hbar^3} \; .
$$
and therefore
\be
v \geq v_{min} = \hbar \; \sqrt3 \; \left(  \frac{\rho}{K \; m^4} \right)^\frac13 
\ee

\medskip

$N$-body simulations \cite{espf1,espf2,espf3,espf4,espf5}
as well as the resolution of Larson's equations \cite{nos2}
point to a cuspy phase-space density behaviour
\be\label{qs}
 Q(r) = Q_h \; \left(\frac{r}{r_h}\right)^{-\beta}
\ee
where $ \beta \simeq 1.9 - 2 , \; r_h $ is the halo radius and $ Q_h $
stands for the mean or characteristic phase space density in the halo.
Classical WDM $N$-body simulations yield results similar to the behaviour
eq.(\ref{qsI}) even for $ r \lesssim  r_h $ when properly using the velocity dispersion of the
keV-scale WDM particles \cite{sinziana,shao}. 

We see that this $ Q(r) $ derived within classical physics tends to infinity for
$ r \to 0 $ violating the Pauli principle bound eq.(\ref{cotQ}).
Therefore, classical physics in galaxies breaks down near the galaxy center.

\medskip

Setting $ \beta =2 $ for simplicity, we find by combining eqs.(\ref{cotQ}) and (\ref{qs})
\be\label{cot}
r \geq r_{min} \equiv \frac{\hbar^{\frac32}}{m^2} \; \sqrt{\frac{Q_h}{K}} \; r_h \; .
\ee
That is, $ r > r_{min} $ given by eq.(\ref{cot}) sets
the domain of validity of classical physics in DM dominated galaxies.

\medskip

The quantum bound eq.(\ref{cotQ}) rules out the presence of galaxy cusps for $ r \lesssim r_{min} $
and implies the existence of galaxy cores with size larger than $ r_{min} $.

\medskip

The values of $ r_h $ and $ Q_h $ vary several orders of magnitude
according to the type of the galaxy (see Table \ref{pgal}). 
$ Q_h $ is larger for dwarf spheroids than for spiral galaxies.
From Table \ref{pgal}
\be\label{databla}
10^{-4} < \frac{\hbar^{\frac32} \; \sqrt{Q_h}}{{\rm (keV)}^2} < 1.3
\ee
The larger value corresponds to ultra compact dwarf spheroidals and the smaller one
to spiral galaxies. We finally have from eqs. (\ref{cot}) and (\ref{databla}),
\be\label{desrq}
\frac{10^{-4}}{\sqrt{K}}  \; \left(\frac{\rm keV}{m}\right)^2 
<   \frac{r_{min}}{r_h} <   \frac{1.3}{\sqrt{K}}  \; \left(\frac{\rm keV}{m}\right)^2 \; .
\ee
$ r_h $ goes from $ \sim 10 $ pc for ultra compact dwarf spheroidals till 
the 10 kpc range for spiral galaxies. 

\medskip
\begin{table}
\begin{tabular}{|c|c|c|c|c|c|} \hline  
 & & & & & \\
 Galaxy  & $ \displaystyle \frac{r_h}{\rm pc} $ & $  \displaystyle \frac{\sigma}{\frac{\rm km}{\rm s}} $
& $ \displaystyle  \frac{\hbar^{\frac32} \;\sqrt{Q_h}}{({\rm keV})^2} $ & 
$ \rho(0)/\displaystyle \frac{M_\odot}{({\rm pc})^3} $ & $ \displaystyle \frac{M_h}{10^6 \; M_\odot} $
\\ & & & & & \\ \hline 
Willman 1 & 33 & $ 4 $ & $ 0.85 $ & $ 6.3 $ & $ 0.39 $
\\ \hline  
 Segue 1 & 38  & $ 4 $ & $ 1.3 $ & $ 10 $ & $ 0.59 $ \\ \hline  
  Leo IV & 151  & $ 3.3 $ & $ 0.2 $ & $ .19 $ & $ 1.14 $ \\ \hline  
Canis Venatici II & 97 & $ 4.6 $ & $ 0.2 $   & $ 0.49 $ & $ 1.43 $
\\ \hline  
Coma-Berenices & 100 & $ 4.6 $  & $ 0.42 $   & 
$ 2.09 $  & $ 1.97 $
\\ \hline  
 Leo II & 233 & $ 6.6 $ & $ 0.093 $  & $ 0.34 $ & $ 7.25 $
\\  \hline  
 Leo T & 152 & $ 7.8 $ &  $ 0.12 $  & $ 0.79 $ & $ 7.4 $
\\ \hline  
 Hercules & 305 & $ 5.1 $ &  $ 0.078 $  & $ 0.1 $ & $ 7.5 $
\\ \hline  
 Carina & 334 & $ 6.4 $ & $ 0.075 $  & $ 0.15 $ & $ 9.6 $
\\ \hline 
 Ursa Major I & 416 & $ 7.6 $ &  $ 0.066 $  & $ 0.25 $ & $ 12.6 $
\\ \hline  
 Draco & 291 & $ 10.1 $ &  $ 0.06 $  & $ 0.5 $ & $ 21 $
\\ \hline  
 Leo I & 388  & $ 9 $ &  $ 0.048 $  & $ 0.22 $ & $ 22 $
\\ \hline  
 Sculptor & 375  & $ 9 $ & $ 0.05 $  & $ 0.25  $ & $ 22.5 $
\\ \hline 
 Bo\"otes I & 322 & $ 9 $ & $ 0.058 $  & $ 0.38 $ & $ 24 $
\\ \hline  
 Canis Venatici I & 750  & $ 7.6 $ & $ 0.037 $ & $ 0.08 $ & $ 27.7 $
\\ \hline  
Sextans & 1019 & $ 7.1 $ & $ 0.021 $ & $ 0.02 $ & $ 35 $
\\ \hline 
 Ursa Minor & 588 & $ 11.5 $ & $ 0.028 $  & $ 0.16 $ & $ 56 $
\\ \hline  
 Fornax  & 944 & $ 10.7 $ & $ 0.016 $  & $ 0.053  $ & $ 74 $
\\  \hline  
 NGC 185  & 355 & $ 31 $ & $ 0.033 $ & $ 4.09 $ & $ 293 $
\\ \hline  
 NGC 855  & 837 & $ 58 $ & $ 0.01 $ & $ 2.64 $ & $ 2480 $
\\ \hline  
  Small Spiral  & 4800  & $ 40.7 $ & $ 0.0018 $ & $ 0.029 $ & $ 5100 $
\\ \hline  
NGC 4478 & 1490 & $ 147 $ & $ 0.003 $ & $ 3.7 $ & $ 1.96   \times 10^4 $
\\ \hline  
 Medium Spiral & $ 1.73 \times 10^4 $ & $ 76.2 $ & $ 3.7 \times 10^{-4} $ & $ 0.0076 $ & $ 6.4 \times 10^4 $
\\ \hline  
 NGC 731 & 4850 & $ 163 $ & $ 9.27 \times 10^{-4} $ & $ 0.47 $ & $ 8.52 \times 10^4 $
\\ \hline 
 NGC 3853   & 4110 & $ 198 $ & $ 8.8 \times 10^{-4} $ & $ 0.77 $  
& $ 8.54 \times 10^4 $ \\ \hline 
NGC 499  & 6070 &  $ 274 $ & $ 5.9 \times 10^{-4} $ & 
$ 0.91 $ & $ 3.27 \times 10^5 $ \\   \hline 
Large Spiral & $ 5.18 \times 10^4 $ & $ 125 $ & $ 0.96 \times 10^{-4} $ & $ 2.3 \times 10^{-3} $ & 
$ 5.2 \times 10^5 $ \\ \hline  
\end{tabular}
\caption{Observed values for $ r_h, \; \sigma, \; \sqrt{Q_h}, \; \rho(0)$ 
and $ M_h $ for galaxies from refs. 
\cite{wp,sltg,dvss,gil,jdsmg,simon11,wolf10,datos1,datos2,martinez}. The phase space density is larger
for smaller galaxies both in mass and size while the surface density $ \mu = \rho(0) \; r_h $ 
is approximately constant \cite{densup1,densup2,densup3,dvss}. Notice that the phase space density is obtained
from the stars velocity dispersion which is expected to be smaller than the DM  velocity dispersion.
Therefore, the reported $ Q_h $ are in fact upper bounds to the true values \cite{jdsmg}.} 
\label{pgal}
\end{table}

The right hand larger value in eq.(\ref{desrq}) corresponds to the smaller galaxies 
(ultra compact dwarfs) with $ r_h $ in the 10 pc range.
The left hand smaller value in eq.(\ref{desrq}) corresponds to the larger galaxies
(spirals) with $ r_h $ in the 10 kpc range. Multiplying both sides of eq.(\ref{desrq}) by $ r_h $
we obtain for a DM particle mass in  the keV scale:
\be\label{rqe}
0.1 \; {\rm pc} \lesssim r_{min}  \lesssim 1 \; {\rm pc} \; ,
\ee
because $ r_h $ is larger for the more dilute galaxies where $ Q_h $ is smaller and vice versa.

\medskip

Notice that eq.(\ref{cot}) provides the {\bf minimal} size of the region where quantum
mechanical effects are strong. 
Quantum mechanical effects will influence a region well beyond
the lower bound $ r_{min} $. Namely, removing the classical cusp which sucks matter 
towards the origin has an effect on the whole galaxy halo.

\medskip

The quantum bound eq.(\ref{pauli}) in the
cosmological context has been  previously considered in refs. \cite{st,gorub,gorubL}
to derive lower bounds on neutral lepton masses.
In the present paper, 
we incorporate to this bound the behaviour of the phase-space density $ Q(r) $  
\cite{espf1,espf2,espf3,espf4,espf5}
to derive lower bounds to the halo radius eq.(\ref{cot}). These bounds are of
semiclassical gravitational nature. 

\bigskip

Assuming, as usual, that virialization holds within the radius $ r_h $, namely
\be\label{vir}
\sigma_h^2 = \frac{G \; M_h}{3 \; r_h}
\ee
where $ M_h $ is the mass within the  radius $ r_h $,
\be\label{mden}
M_h  = \frac43 \; \pi \; r_h^3 \; \rho_h  \quad , \quad Q_h = \frac{\rho_h}{\sigma_h^3} \; ,
\ee
and  $ G $ is Newton's gravitational constant, we obtain from eqs. 
(\ref{cot}), (\ref{vir}) and (\ref{mden}),
\be
Q_h = \frac9{4 \, \pi} \; \sqrt{\frac3{M_h \; (G \; r_h)^3}} \; ,
\ee
and
\be\label{mpla}
 r_{min} =  \frac3{2 \; \sqrt{\pi \; K} \;  m^2} \;
\left(\frac{\hbar}{\sqrt{G}}\right)^{\frac32} \;
\left( \frac{3 \; r_h}{M_h}\right)^{\frac14} = \frac{3 \; \hbar}{2 \; \sqrt{\pi \; K}}
\; \left( \frac{m_{Pl}}{m} \right)^2 \; \left( \frac{3 \; r_h}{G \; M_h}\right)^{\frac14}
 \quad , 
\ee
where $ m_{Pl} = \sqrt{\hbar/G} $ is Planck's mass. More explicitly
\be \label{rqf}
\frac{r_{min}}{\rm pc} = \frac{0.58793}{\sqrt{K}} \; \left(\frac{r_h}{\rm pc} \;  
\frac{10^6 \; M_\odot}{M_h}\right)^{\frac14} \;  \left(\frac{\rm keV}{m}\right)^2 \; ,
\ee
where we used that
$$
\left(\frac32\right)^{\frac54}  \;  \frac1{\sqrt{\pi} \; 5^{\frac14}} \; \;
\frac{\hbar^{\frac32}}{({\rm kpc})^{\frac34} \;  M_\odot^{\frac14}
\; G^{\frac34} \; ({\rm keV})^2} = 0.58793 \ldots  \; .  
$$
The values of the constant $ 0.60364/\sqrt{K} $ are displayed in table \ref{valK}
while those of $ r_h/{\rm pc} $ and $ 10^{-6}M_h/M_\odot $ are  displayed in table
\ref{pgal}.

\medskip

It must be stressed that $ r_{min} $ is the {\bf minimal} value for the core radius.
The core radius can be {\bf well above} this lower bound which corresponds
to maximally packed fermions around the center of the galaxy.
Hence, one expects for diluted objects as galaxies core radii much larger than 
the lower bound $ r_{min} $.
Even in atoms the phase-space density turns to be significantly below the quantum bound eq.(\ref{cotQ}) 
for thermal fermions \cite{qm}. Moreover, our derivation of $ r_h $ within the Thomas-Fermi
approach presented in sec. \ref{tf} shows that $ r_h \gg r_{min} $ for dwarf galaxies.

\bigskip

The quantum bound eq.(\ref{cotQ}) applies 
for any kind of  fermionic dark matter. However, the value of
$ r_{min} $ strongly depends on the DM particle mass.

\medskip

For CDM where 1 GeV $ < m < $ 100 GeV,
we see that $ r_{min} $ can be from eq.(\ref{rqf}) between dozens of kilometers and a few meters.
That is, $ r_{min} $ for CDM is astronomically compatible with zero.
Therefore, classical aproaches as  $N$-body simulations and the classical
Boltzmann-Vlasov equation fully apply for CDM and unavoidably produce cusps.

\medskip

For HDM where  1 eV $ < m < $ 10 eV, we find $ r_{min} $ from eq.(\ref{rqf}) between the kpc
and the Mpc, depending on the galaxy type. 
HDM suppresses all structures scales below the Mpc scale due to its long
free-streaming length and has been ruled out years ago.
Anyway, it is important to learn that  classical aproaches to HDM as
$N$-body simulations are not valid for scales below $ r_{min} $ which
can be from the kpc to the Mpc scale according to the type of galaxy
and the value of $ m $ in the eV scale. 

\medskip

For bosonic DM, the bound eq.(\ref{cotQ}) does not apply
and the formation of cusps is allowed. However, the observed galaxy density profiles 
are cored \cite{obs1,obs2,obs3,obs4,gil,wp} (see also \cite{sg}).

\medskip

In all cases cusps of fermionic DM
in the galaxy density profile are artifacts 
produced by classical physics computations irrespective of the nature of dark matter
(HDM, WDM, CDM) and of whether the computations are numerical or analytical.

That is, quantum physics, namely
the Pauli principle, {\bf rule out} galaxy cusps for fermionic dark matter.

\medskip

We have so far ignored baryons in our analysis of the galaxies.
This is fully justified for dwarf spheroidal galaxies which are composed today
of 99.99\% of dark matter \cite{barena1,barena2,datos1,datos2,mwal}. In large galaxies the baryon fraction can
reach values up to  1 - 3 \% \cite{bariones1,bariones2,bariones3}.
We have also ignored supermassive central black holes which
appear in large spiral galaxies but have not been observed in dwarf galaxies.
As it is known, the formation of supermassive central black holes is correlated
to the formation of the galaxy itself but this whole issue is beyond the scope of the 
present paper.
Anyway, it must be noticed that the central black hole mass is at most 
$ \sim 10^{-3} $ of the mass of the bulge.
Baryons can be important in the galaxy formation and evolution, a subject
that is outside the scope of our paper.
Fermionic WDM by itself produces galaxies and structures in agreement with observations.
Therefore, the effect of including baryons is expected to be a correction to the pure WDM results
presented in this paper, consistent with the fact that dark matter is in average six times more abundant than baryons.

\subsection{Classical upper bounds from the DM phase-space density evolution}

The quantum upper bound on the phase-space density eq.(\ref{cotQ}) 
is unrelated to the classical upper bound derived from the self-gravity classical dynamics
as already noticed in ref. \cite{st}.
Namely, the fact that the coarse-grained phase-space density when
  averaged also in $\vec{r}$--space over a galaxy halo,
can only {\bf decrease} by classical collisionless phase mixing or classical  self-gravity dynamics
\cite{psd1,psd2,psd3,psd4,psd5,bdvs2,dvs}. This averaged phase-space density $ \bar Q $ is thus bounded 
by its primordial value $ Q_{prim} $.

This upper bound follows from the {\bf classical}
self-gravity dynamics evolution contrary to  eq.(\ref{cotQ}) which applies
to fermionic DM and has a quantum origin on the Pauli principle.
In addition, this classical upper bound only applies to the phase-space density
averaged over a large volume and on time intervals: $ {\bar Q} < Q_{prim} $.

\medskip

At specific points like the galaxy center the phase-space density
is classically unbounded due to the appearence of cusps and 
certainly much larger than the primordial phase-space density.

The primordial DM phase-space density which is space-independent can be written as \cite{dvs}
\be\label{qprim}
Q_{prim} = \frac{\sqrt{27} \; g_i}{2 \; \pi^2 \; \hbar^3} \; m^4 \; \frac{I_2^{\frac52}}{I_4^{\frac32}}\; , 
\ee
where $ g_i $ is the number of internal degrees of freedom of the DM particle
($ g_i = 4 $ for Dirac fermions), 
$ I_2 $ and $ I_4 $ are the dimensionless momenta of the primordial DM distribution function.
Using eq.(\ref{qprim}) for $ Q_{prim} $, its  self-gravity dynamical evolution and observational data
for the phase-space density today (as in Table \ref{pgal}) indicated  that
the DM particle mass is in the keV scale \cite{dvs}.

\medskip

Inserting eq.(\ref{qprim}) into the quantum bound eq.(\ref{cotQ}) yields
$$
\sqrt{27} \; I_2^{\frac52} \leq K \; \pi^2 \; I_4^{\frac32} \; .
$$  
It is easy to check that this inequality is fulfilled for the distribution functions
considered in the appendix \ref{calK} as it must be.

That is, the primordial value $ Q_{prim} $ of the phase-space density safely
fulfills the quantum  bound eq.(\ref{cotQ}).

\section{Gravitation and the WDM pressure: the main drivers of galaxy formation}

As is well known, galaxy formation as all structure formation in the Universe
is driven by gravitational physics.

The main notion regarding structure formation at a given scale $ r $ is the
Jeans' length $ \lambda_J(r) $ at this scale. The Jeans' length is the product of the speed of sound
or the velocity of the particles $ v $ times the free fall time
\be \label{lj}
\lambda_J(r) = v(r) \; \sqrt{\frac{\pi}{G \; \rho(r)}}
\ee
where $ G $ is Newton's gravitational constant. 

Length scales larger than  $ \lambda_J(r) $ are gravitationally unstable. Therefore,
$ \lambda_J(r) $ indicates the scale of the largest objects that can be formed
at the distance $ r $. The mass density $ \rho(r) $ decreases from its maximum value at
the center of the galaxy $ r = 0 $ till 
asymptotic typical values 200 times the average density in the Universe
at $ R_{200} \sim R_{virial} $.

At a scale $ r $ structures larger than $ \lambda_J(r) $ can not form because
they are unstable while structures smaller than $ \lambda_J(r) $ can grow
reaching a size $ \sim r $. Therefore,
$$
 r \sim \lambda_J(r) \; .
$$

\medskip

During the linear regime of the cosmological evolution 
the phase space density is small enough and the  Pauli principle is always fulfilled.
Notice that the phase space measure $ d^3 p \; d^3 r $ is invariant under
the universe expansion and therefore the  universe expansion by itself
does not affect the fullfiling of the Pauli principle.

\medskip

Only when structure formation increases the particle density and 
particle velocity, the phase space distribution function $ f({\vec r},{\vec p}) $ 
may increase to approach
its quantum upper bound $ 2 $ for spin one-half fermions.

\medskip

In summary, starting from the linear cosmological fluctuations at WDM decoupling, 
we can classically evolve the phase space distribution function 
$ f({\vec r},{\vec p}) $ as long as $ f $ remains well below its quantum upper bound.
As soon as classical physics breaks in some region, even small, we have to use a quantum mechanical
treatment not only in the region where a classical $ f $ would violate the quantum upper bound
but in the whole region around. This is not an easy problem and may be 
solved using quantum Monte Carlo methods,
time dependent Hartree-Fock methods and quantum Boltzmann equations
evolution \cite{tdhf1,tdhf2,tdhf3,tdhf4,tdhf5}. 

Clearly, one can use the classical evolution in most of the space where
 $ f $ stays well below the quantum bound and match it with the quantum evolution
around the galaxy centers. 

\medskip 

The Jeans' mass is given by
\be \label{mj}
M_J = \frac43 \; \pi \; \rho \; \lambda_J^3 =  \frac43 \; \pi \; v^3 \; 
\sqrt{\left(\frac{\pi}{G}\right)^3 \; \frac1{\rho}} \; . 
\ee
In terms of the phase-space density, $ \lambda_J $ and $ M_J $ read
$$
\lambda_J = 3^\frac34 \; \sqrt{\frac{\pi}{G \; v \; Q}} \quad , \quad
M_J = \frac{4 \; \pi^\frac52}{3^\frac14} \; \sqrt{\frac{v^3}{G^3 \; Q}}  \; .
$$
Then, from eq.(\ref{cotI}) and eq.(\ref{qsI}) with $ \beta = 2 $ we have the bound
\be
\lambda_J(r) \geq \sqrt{\frac{3 \, \pi}{K \; \sigma(r)}} \; 
\sqrt{\frac{\hbar}{G}} \; \frac{\hbar}{ m^2} \; ,
\ee
More explicitly, we get a lower bound of the product of the Jeans' length and
the square root of the velocity dispersion
$$
\sqrt{\sigma(r)} \; \; \lambda_J(r)  \geq \frac{0.339}{\sqrt{K}} \; \left(\frac{\rm keV}{m}\right)^2 \; \; {\rm pc}
$$
showing that neither the  Jeans' length nor the velocity dispersion can vanish for $ r \to 0 $.

\subsection{Dwarf galaxies as WDM quantum macroscopic objects}\label{broglie}

To determine whether a system of particles has a classical or quantum nature
we should compare the particle de Broglie wavelength with the interparticle distance.

\medskip

The de Broglie wavelength of DM particles in a galaxy can be estimated as
\be\label{LdB}
\lambda_{dB}  = \frac{\hbar}{m \; \sigma} \; ,
\ee
while the average interparticle distance $ d $ can be estimated as
\be\label{dis}
d = \left( \frac{m}{\rho_h} \right)^\frac13 \; ,
\ee
where $ \rho_h $ is the average density in the  galaxy core.
By using $ \rho_h = \sigma^3 \; Q_h $ and eqs.(\ref{LdB})-(\ref{dis}), 
we can express the ratio
$$ 
{\cal R} \equiv \frac{\lambda_{dB}}{d}
$$
as, 
\be
{\cal R} = \hbar \; \left( \frac{Q_h}{m^4}\right)^\frac13 \; .
\ee

Using now the observed  values of $ Q_h $ from Table \ref{pgal} yields
\be
2 \times 10^{-3}  < {\cal R} \; \left( \frac{m}{\rm keV}\right)^\frac43 < 1.4
\ee
The larger value of $ \cal R $ is for ultracompact dwarfs and the smaller value of $ \cal R $ 
is for big spirals.

\medskip

The ratio $ \cal R $ around unity clearly implies a macroscopic quantum object.

Notice that here as well as in the bound eq.(\ref{cotQ}) $ \hbar^3 \; Q/m^4 $ measures how quantum 
or classical is the system (the galaxy). 

\medskip

We conclude {\bf solely from observations} 
that compact dwarf galaxies are natural macroscopic quantum objects for WDM.

\subsection{Dwarf Galaxies supported by WDM fermionic quantum pressure}\label{qup}

For an order--of--magnitude estimate, let us consider a halo of mass $ M $ and radius 
$ R $ of fermionic matter.
It can be fermionic DM or baryons. Each fermion can be considered inside
a  cell of size $ \Delta x \sim 1 / n^{\frac13} $ and therefore has a momentum
$$
p \sim \frac{\hbar}{\Delta x} \sim \hbar \; n^{\frac13} \; .
$$
The associated quantum pressure $ P_q $ (flux of the momentum) has the value 
\be\label{presq}
P_q = n \;\sigma\; p \sim \hbar \;\sigma\; n^{\frac43} = \frac{\hbar^2}{m} \; n^{\frac53} \; .
\ee
where $ \sigma $ is the mean velocity given by
$$
\sigma = \frac{p}{m} = \frac{\hbar}{m} \; n^{\frac13} \; .
$$ 
The system will be in dynamical equilibrium if this quantum pressure is balanced by
the gravitational pressure
\be\label{pgr}
P_G = {\rm gravitational~ force}/{\rm area} = \frac{G \; M^2}{R^2} \times \frac1{4 \, \pi \; R^2}
\ee
We estimate the number density as
$$
n = \frac{M}{\frac43 \; \pi \; R^3 \; m} \; ,
$$
and we use that $ p = m \;\sigma$ to obtain from eq.(\ref{presq})
\be\label{pqu}
 P_q = \frac{\hbar^2}{m \; R^5} \; \left(\frac{3 \; M}{4 \; \pi \; m}\right)^\frac53 \; .
\ee
Equating $  P_q = P_G $ from eqs.(\ref{pgr})-(\ref{pqu})
yields the following relations between the size $ R $ and the velocity $\sigma$ 
with the mass $ M $ of the system:
\bea\label{estM}
&& R = \frac{3^\frac53}{(4 \; \pi)^\frac23} \;
\frac{\hbar^2}{G \; m^\frac83 \; M^\frac13} = 10.6 \ldots {\rm pc} \;
\left(\frac{10^6 \;  M_\odot}{ M}\right)^\frac13 \; \left(\frac{\rm keV}{m}\right)^\frac83 \\ \cr
&&\sigma= \left(\frac{4 \, \pi}{81}\right)^\frac13 \; \frac{G}{\hbar} \;  m^\frac43 \; M^\frac23=
22.9 \ldots \frac{\rm km}{\rm s} \; \left(\frac{m}{\rm keV}\right)^\frac43 \; 
\left(\frac{M}{10^6 \;  M_\odot}\right)^\frac23 \;   .
\eea
Notice that the values of $ M , \;  R $ and $ \sigma $ are consistent with
dwarf galaxies. Namely, for $ M $ of the order $ 10^6 \;  M_\odot $
(typical mass value for dwarf spheroidal galaxies), $ R $ and  $\sigma$ 
have the correct order of magnitude for dwarf spheroidal galaxies 
for a WDM particles mass in the keV scale (see Table \ref{pgal}).

These results back the idea that dwarf spheroidal galaxies are supported by the
fermionic {\it WDM quantum pressure} eq.(\ref{pqu}).

\medskip

It is useful to express the above quantities in terms of the density $ \rho $, as
follows
\bea\label{semic}
&& M = \frac{9 \; \hbar^3}{2 \; m^4} \; \sqrt{\frac{\rho}{\pi \; G}} =  0.7073 \ldots 10^5 \; M_\odot \;
\sqrt{\rho \; \frac{{\rm pc}^3}{ M_\odot}} \; \left(\frac{\rm keV}{m}\right)^4  \; ,
\\ \cr
&& R = \frac{3 \; \hbar}{2 \; \sqrt{\pi \; G}} \; \frac1{m^\frac43 \; \rho^\frac16} = 
31.05 \ldots \; {\rm pc} \; \left(\frac{M_\odot}{\rho \; {\rm pc}^3 }\right)^{\frac16} 
\; \left(\frac{\rm keV}{m}\right)^{\frac43} \; ,
\cr \cr
&&\sigma= \hbar \; \left( \frac{\rho}{m^4}  \right)^\frac13 = 1.990 \ldots \; \frac{{\rm km}}{{\rm s}}
\; \left( \rho \; \displaystyle \frac{{\rm pc}^3 }{M_\odot}\right)^{\frac13} \;
\left(\frac{\rm keV}{m}\right)^{\frac43}  \; ,\label{vsemic}\\ \cr
&& P_q = \hbar^2 \; \frac{\rho^\frac53}{m^\frac83}  = 4.399 \; 10^{-11} \; \frac{M_\odot}{{\rm pc}^3 }
\; \left( \rho \; \displaystyle \frac{{\rm pc}^3 }{M_\odot}\right)^{\frac53} \;
\; \left(\frac{\rm keV}{m}\right)^{\frac83} \; ,
\eea
$ R $ and $ M $ are typical {\it semiclassical gravitational} quantities involving both $ G $ and
$ \hbar $. The particle velocity $ \sigma $ and the pressure $ P_q $ are of purely quantum 
mechanical origin.

The radius $ R $ and mass $ M $ are the {\it semiclassical} Jeans' length and Jeans'
mass as can be seen by inserting the velocity from eq.(\ref{vsemic}) in eqs.(\ref{lj}) 
and (\ref{mj}) with the result
$$
\lambda_J = \frac{2 \; \pi}3 \; R \quad , \quad M_J = \left(\frac{2 \; \pi}3\right)^3 \; M
$$
The associated phase-space density results
$$
Q = \frac{m^4}{\hbar^3} \; .
$$
In this case where $ v $ has purely quantum origin [eq.(\ref{vsemic})], the de Broglie wavelength 
is equal to the interparticle distance $ \lambda_{dB} = d $. Thus the ratio $ {\cal R} $ is unity
which is the extreme (fermion degenerate) quantum case.

\medskip

In the presence of squared angular momentum $ L^2 $ we have to add the centrifugal pressure
$$
P_L = \frac1{4 \, \pi \; R^2} \; \frac{L^2}{M \; R^3}
$$
in the equilibration equation $ P_q + P_L = P_G $. The relation between the radius, mass
and velocity eq.(\ref{estM}) takes now the form
\be
R =  \frac{L^2}{G \; M^3} +\frac{3^\frac53}{(4 \; \pi)^\frac23} \;
\frac{\hbar^2}{G \; m^\frac83 \; M^\frac13} \; .
\ee
A simple estimation of the total angular momentum of the whole halo goes as follows
$$
L^2 \sim  \frac12 \; M^2 \; R^2 \;  3 \; \sigma^2
$$
where the $ \frac12 $ factor comes from averaging the $ \sin^2 $ of the angle between 
the momentum $ \vec p $ and the particle position $ \vec r $, and the factor $ 3 $ comes 
from the relation $ v^2 = 3 \; \sigma^2 $. We thus obtain 
\be
R =10.6 \ldots {\rm pc} \left(\frac{10^6 \;  M_\odot}{ M}\right)^\frac13 \; 
\left(\frac{\rm keV}{m}\right)^\frac83 +
3.48 \ldots {\rm pc} \; \frac{10^6 \;  M_\odot}{ M} \;
\left(\frac{\sigma}{10 \; \frac{\rm km}{\rm s}} \; \frac{R}{10 \; {\rm pc}}\right)^2 \; .
\ee
We see that the angular momentum contribution increases the size $ R $.
However, for dwarf galaxies, we see that $ R $ and $\sigma$ have the same order of magnitude 
for $ L > 0 $ and for $ L = 0 $.

\section{Galaxy properties from quantum  fermionic WDM in the Thomas-Fermi approach}\label{tf}

DM particles are nonrelativistic during structure formation and their
chemical potential is given by 
\be\label{potq}
\mu(r) =  \mu_0 - m \; \phi(r) 
\ee
where $ m $ is the mass of the DM particle, $ \mu_0 $ is a constant and 
$ \phi(r) $ is the gravitational potential.

\medskip

We consider for simplicity the spherical symmetric case where
the Poisson equation for the gravitational potential takes the form
\be\label{pois}
\frac{d^2 \mu}{dr^2} + \frac2{r} \; \frac{d \mu}{dr} = - 4 \, \pi \; G \; m \; \rho(r)
\ee
where $ G $ is Newton's gravitational constant and $ \rho(r) $ is the DM mass density.

\medskip

Since the DM mass density is bounded by the Pauli principle as analyzed in sec. \ref{cotaq},
$ \rho(r) $ must be bounded at the origin and therefore we must impose 
as boundary condition at the origin:
\be
\frac{d \mu}{dr}(0) = 0 \; .
\ee
We can write the DM mass density as the integral over the momentum of the DM distribution
function
\be\label{den}
\rho(r) = \frac{g \; m}{2 \; \pi^2 \; \hbar^3} \int_0^{\infty} p^2 \; dp \; f[e(p)-\mu(r)]
\ee
where $ e(p) = p^2/(2 \, m) $ is the particle kinetic energy, $ f(E) $ is the
energy distribution function and
$ g $ is the number of internal degrees of freedom of the DM particle.
$ g = 1 $ for Majorana fermions and $ g = 2 $ for Dirac fermions. 

\medskip

Eqs.(\ref{pois}) and (\ref{den}) provide a system of ordinary nonlinear differential
equations that determine 
the chemical potential $ \mu(r) $ and constitutes the Thomas-Fermi semi-classical approximation.
Fermionic DM in the Thomas-Fermi approximation has been previously considered
in ref. \cite{peter4,peter1,peter3,peter2}.

The Thomas-Fermi equations provide a semi-classical approximation
in which the energy distribution function $ f(E) $ is given.

\medskip

The boundary condition for the chemical potential is usually set to zero at a given radius $ R $
which is the appropriate condition for the Coulomb interaction in atoms \cite{qm} and for 
selfgravitating degenerate fermions \cite{ll}. In the selfgravitating degenerate case the mass density
is proportional to the power $ 3/2 $ of the chemical potential and therefore both quantities vanish
at the same point.  In the general selfgravitating case for
non-degenerate fermions where eq.(\ref{den}) applies, the  
mass density stays nonzero beyond the point where the chemical potential changes sign 
and becomes negative.

\medskip

We integrate the Thomas-Fermi nonlinear differential
equations (\ref{pois})-(\ref{den}) from $ r = 0 $ till the 
boundary $ r = R = R _{200} \sim R_{vir} $ defined as the radius where the 
mass density equals $ 200 $ times the mean DM density.

\bigskip

In order to determine the distribution function $ f(E) $ one should follow the precise DM evolution
since decoupling. Such evolution must take into account the quantum character
that emerges when the distribution function approaches the quantum upper bound eq.(\ref{cotI}).
This quantum dynamical calculation is beyond the scope of the present paper.

We modelize here the distribution function $ f(E) $ by equilibrium
Fermi-Dirac functions and by out of equilibrium Dodelson-Widrow and $ \nu$-MSM 
distribution functions (see the appendix \ref{calK}).

\bigskip

Let us define dimensionless variables $ \nu(\xi) , \;  \xi , \; \xi_0 $ 
and the dimensionless distribution  function $ \Psi $ as
\be\label{varsd}
r = L_0 \; \xi \qquad , \qquad R = L_0 \; \xi_0  \qquad , \qquad 
\mu(r) =  E_0 \;  \nu(\xi) \qquad , \qquad f(E) = \Psi \! \left[\frac{E}{ E_0}\right]  \; ,
\ee
where $ E_0 $ is the characteristic comoving energy of the DM particles at decoupling.
The characteristic length $ L_0 $ emerges from the dynamical equations 
(\ref{pois})-(\ref{den}) and is given by 
\be\label{defsig}
L_0 \equiv \frac{\sqrt{3 \, \pi \; \hbar^3}}{\sqrt{G} \; (2\,m)^2}  \; 
\left(\frac{2 \, m}{ E_0}\right)^{\frac14}  \; ,
\ee

\medskip

For $ g = 2 $ the mass density takes the form
\be\label{den2}
\rho(r) = \frac{m^4}{\pi^2 \; \hbar^3} \; \left(\frac{2 \,  E_0}{m}\right)^{\frac32} \; \beta(\nu(\xi))
\quad , \quad \beta(\nu) \equiv \int_0^{\infty} y^2 \; dy \; \Psi(y^2 -\nu) \; ,
\ee
where we use the integration variable $ y \equiv p / \sqrt{2 \, m \;  E_0} $
and the Poisson equation (\ref{pois})-(\ref{den}) become
\be\label{nu}
\frac{d^2 \nu}{d\xi^2} + \frac2{\xi} \; \frac{d \nu}{d\xi} = -3 \; \beta(\nu(\xi))
\quad ,  \quad \rho(\xi_0) = 200 \; {\bar \rho}_{DM} \quad ,  \quad
\nu'(0) = 0 \quad  , 
\ee
The distribution functions $ \Psi $ are given in the Appendix \ref{calK} for the thermal 
and out of equilibrium cases.

\medskip

In order to integrate eq.(\ref{nu}) we have to specify $ \nu(0) $. $ \nu(0) $ is determined by 
the value of the phase space density at the origin $ Q(r=0) $. That is, to find a given galaxy 
we must give $ Q(r=0) $ and the boundary radius $ R $ in dimensionless variables $ \xi_0 $. 
In eq.(\ref{Qnu0}) below we find the relation between $ \nu(0) $ and $ Q(r=0) $
which permits to compute $ \nu(0) $ from a given value of $ Q(r=0) $.

\subsection{Main physical galaxy properties: mass, velocity dispersion, density and pressure}

We derive in this subsection the expressions for the main physical properties of the galaxies 
(mass, velocity dispersion, density and pressure) in the Thomas-Fermi semi-classical approximation.

\medskip

The average velocity of the particles is space-dependent and follows
from the average momentum given by
\be\label{velo}
v^2(r) = \frac1{m^2} \; 
\frac{\int_0^{\infty} p^4 \; dp \; f[e(p)-\mu(r)]}{\int_0^{\infty} p^2 \; dp \; f[e(p)-\mu(r)]}
= \frac{2 \,  E_0}{m} \; \alpha^2(\nu(\xi)) \; ,
\ee
where
\be\label{defalf}
\alpha(\nu)  \equiv \sqrt{\frac{\int_0^{\infty} y^4 \; dy \; \Psi(y^2 - \nu)}{\int_0^{\infty} 
y^2 \; dy \; \Psi(y^2 - \nu)}} \; .
\ee

\bigskip

The mass enclosed in the sphere of radius $ R $ follows by integrating the mass density
given by eq.(\ref{den2})
\be\label{ma1}
M = 4 \, \pi \int_0^R r^2 \; dr \; \rho(r) = \frac{\sqrt{27 \, \pi}}{2 \; m^2} \;
\left(\frac{\hbar}{G}\right)^{\frac32} \; \left(\frac{ E_0}{2 \, m}\right)^\frac34 \; 
\int_0^{\xi_0} \xi^2 \; d\xi \; \beta(\nu(\xi)) \; ,
\ee
where we used eqs.(\ref{varsd}) and (\ref{defsig}).
The integral over $ \xi $ can be performed with the help of eq.(\ref{nu}),
\be\label{ma2}
\int_0^{\xi_0} \xi^2 \; d\xi \; \beta(\nu(\xi)) = - \frac13 \; \xi_0^2 \; \nu'(\xi_0) \; .
\ee
Notice that $ \nu'(\xi_0) $ is always negative since the function $ \beta(\nu) $ is positive definite.

\medskip

We can write from eqs.(\ref{ma1})-(\ref{ma2}) the mass $ M $ as 
\be\label{ma3}
M =  \frac{\sqrt{3 \, \pi}}{2 \; m^2} \; \left(\frac{\hbar}{G}\right)^{\frac32} \;
\left(\frac{ E_0}{2 \, m}\right)^\frac34
\; \xi_0^2 \;  \left| \nu'(\xi_0) \right| \; .
\ee

\medskip

The pressure at the point $ r , \; P(r) $ is given by an analogous integral \cite{bdvs2}
\be\label{pres}
P(r) = \frac{m^4}{3 \, \pi^2 \; \hbar^3} \; \left(\frac{2 \,  E_0}{m}\right)^\frac52 
\; \int_0^{\infty} y^4 \; dy \; \Psi[y^2 - \nu(\xi)] = \frac{m^4}{3 \, \pi^2 \; \hbar^3} \; 
\left(\frac{2 \,  E_0}{m}\right)^\frac52 \; \alpha^2(\nu(\xi)) \; \beta(\nu(\xi)) \; .
\ee
From eqs.(\ref{den2}) and (\ref{velo}) we can write the density and the average square velocity as
$$
\rho(r) = \rho(0) \; \frac{\beta(\nu(\xi))}{\beta_0} \quad , \quad
v(r) =v(0) \;  \frac{\alpha(\nu(\xi))}{\alpha_0} \; ,
$$
Here $ \alpha_0 \equiv  \alpha(\nu(0)) $ and $ \beta_0 \equiv  \beta(\nu(0)) $.

Moreover, from eqs.(\ref{den2}), (\ref{velo}) and (\ref{pres}) we derive as local
equation of state:
\be\label{eces}
P(r) =  \frac13 \; v^2(r) \; \rho(r) \;  \; .
\ee
This local equation of state generalizes the local perfect fluid equation of state 
for $r$-dependent velocity $ v(r) $.
As we see below, the perfect fluid equation of state
is recovered both in the classical dilute limit and in the quantum degenerate limit.

It is very instructive to compute the derivative of the pressure eq.(\ref{pres}) with respect to 
$ r = L_0 \; \xi $. Upon integrating by parts and using eqs.(\ref{potq}), 
(\ref{varsd}) and (\ref{den2}) we find
\be\label{ehidr}
\frac{dP}{dr} + \rho(r) \; \frac{d\phi}{dr} = 0 \; .
\ee
This shows that the quantum Thomas-Fermi equation implies the hydrostatic equilibrium equation.

\medskip

By analogy with the Burkert density profile, 
the halo radius $ r_h = L_0 \; \xi_h $ in this theoretical calculation can be defined in
dimensionless variables as,
$$
\frac{\rho(r_h)}{\rho(0)} =\frac{\beta(\xi_h)}{\beta_0} = \frac14 \; .
$$
Furthermore, from eqs.(\ref{den2}) and  (\ref{velo}) the phase space density $ Q(r) $ 
is given by
\be\label{Qab}
Q(r) = \frac{\sqrt{27}}{\pi^2 \; \hbar^3} \; \; m^4 \; \; \frac{\beta(\xi)}{ \alpha^3(\xi) } 
\quad {\rm and} \quad Q(0) = \frac{\sqrt{27}}{\pi^2 \; \hbar^3} \; \; m^4 \; \; 
\frac{\beta_0}{\alpha_0^3} \; ,
\ee
which turns to be independent of $ E_0 $. 

\medskip

From eqs.(\ref{varsd}), (\ref{defsig}) and (\ref{ma3}), 
$ R^3 \; M $ turns to be independent of $ E_0 $ too:
$$
R^3 \; M = \frac{(3 \; \pi)^2 \; \hbar^6}{G^3 \; (2 \; m)^8} \; \;  
\xi_0^5 \;  \left| \nu'(\xi_0) \right| \; .
$$

We expressed above the main physical galaxy magnitudes
$ L_0, \; M, \; v(r) , \; P(r) $ and $ \rho(r) $ in terms of the DM characteristic energy
$ E_0 $ and the potential $ \nu(\xi) $, solution of eq.(\ref{nu}).
Because $ E_0 $ is not directly observed we will eliminate  
$ E_0 $ in terms of  $ \rho(r=0) $ from eq.(\ref{den2}). [We may also choose other point $ r \neq 0 $]. 
We obtain
\be\label{Txi}
\left(\frac{2 \,  E_0}{m}\right)^{\frac32} =  \frac{\pi^2 \; \hbar^3}{m^4} \; 
\frac{\rho(0)}{\beta_0} \; .
\ee
By eliminating $ E_0 $ and using eqs.(\ref{den2})-(\ref{defalf}), we have for the pressure eq.(\ref{pres})
\be\label{pres2}
P(r) = \frac{\pi^{\frac43}}3 \; \hbar^2 \; m^4 \; \left[\frac{\rho(0)}{\beta_0 \; m^4}\right]^{\frac53}
\; \alpha^2(\nu) \; \beta(\nu)  \quad {\rm and} \quad P(0) = \frac{\pi^{\frac43}}3 \; \hbar^2 \; m^4 \;
\left[\frac{\rho(0)}{m^4}\right]^{\frac53} \; \left(\frac{\alpha_0}{\beta_0^{\frac13}}\right)^2\; .
\ee
Inserting  eq.(\ref{Txi}) in eqs.(\ref{varsd}), (\ref{defsig}), 
(\ref{velo}) and (\ref{ma3}) yields
\be
L_0 = \sqrt{\frac3{8 \, G}} \;  \frac{\hbar}{m^{\frac43}} \; 
\left[\frac{\pi \; \beta_0}{\rho(0)}\right]^{\frac16} \quad , \quad
v(r) = \hbar \; \alpha(\nu) \; \left[\frac{\pi^2 \; \rho(0)}{\beta_0 \; m^4}\right]^{\frac13}
\quad , \quad 
M =  \frac{\sqrt3 \; \pi^{\frac32} \; \hbar^3}{2^{\frac52} \; G^\frac32 \; m^2}  \; 
\; \; \xi_0^2 \; \left| \nu'(\xi_0) \right| \sqrt{\frac{\rho(0)}{\beta_0 \; m^4}} \; .
\ee

\medskip

More explicitly, we can express $ L_0 , \; M $ and $ v(r) $ as
\bea\label{l0}\label{eregc}
&& L_0 = R_0 \; \; \left(\frac{\rm keV}{m}\right)^{\frac43} \; 
\left[\frac{\beta_0 \; M_\odot}{\rho(0) \; {\rm pc}^3 }\right]^{\frac16} 
\quad , \quad  R_0 \equiv \sqrt{\frac3{8 \, G}} \;  \frac{\hbar}{{\rm keV}^{\frac43}} \;
 \left[\frac{\pi \; {\rm pc}^3}{M_\odot}\right]^{\frac16} = 22.47 \; {\rm pc} \; , \label{rgalc}
\\ \cr \cr
&& M = M_0 \; \; \sqrt{\rho(0) \; \frac{{\rm pc}^3}{ M_\odot}} \; \; \left(\frac{\rm keV}{m}\right)^4
\; \; \frac{\xi_0^2}{\sqrt{\beta_0}} \; \; \left| \nu'(\xi_0) \right|
\quad , \quad  M_0 \equiv \frac{\sqrt3 \; \pi^{\frac32} \; \hbar^3}{2^{\frac52} \; 
G^\frac32 \; {\rm keV}^4}  \; \sqrt{\displaystyle\frac{M_\odot}{{\rm pc}^3}} = 
1.425 \; 10^5 \; M_\odot \; ,\label{masgc} \cr \cr \cr
&&  v(r) = v_0 \; \; 
\left[ \rho(0) \; \displaystyle \frac{{\rm pc}^3 }{M_\odot}\right]^{\frac13} \;
\left(\frac{\rm keV}{m}\right)^{\frac43} \; \; \frac{\alpha(\nu)}{\beta_0^{\frac13}}
\quad , \quad v_0 \equiv \pi^{\frac23} \;
\left[\displaystyle \frac{M_\odot}{{\rm pc}^3}/ {\rm keV}^4 \right]^{\frac13} 
= 4.268 \; \; \frac{{\rm km}}{{\rm s}} \; .\label{velgc}
\eea
The {\it semiclassical} galaxy magnitudes $ L_0, \; M, \;  v(0) $ and $ P(0) $ emerging from the
Thomas-Fermi equations generalize the corresponding expressions eq.(\ref{semic})-(\ref{vsemic}) 
derived just equating the gravitational and WDM quantum pressures and describing the quantum
degenerate fermions limit. Eqs.(\ref{l0})-(\ref{velgc}) cover the full range of physical situations
from the degenerate fermions till the 
dilute classical limit as we discuss in the next subsection.

\medskip

The local Jeans' length and Jeans' mass follow by inserting eqs.(\ref{den2})  and (\ref{velo})
for $ \rho(r) $ and $ v(r) $, respectively, into eqs.(\ref{lj}) and (\ref{mj}) 
\be
\lambda_J(\xi) = \pi \; \sqrt{\frac83} \; L_0 \; 
\frac{\alpha(\xi)}{\sqrt{\beta(\xi)}} \quad , \quad
M_J(\xi) = \frac{8 \, \pi^3}{\sqrt{27}} \; \; M \; \;  \frac{\alpha^3(\xi)}{\sqrt{\beta(\xi)}} 
\frac1{\xi_0^2} \; \left| \nu'(\xi_0) \right|
\ee
We see that $ \lambda_J $ and $ M_J $ differ from $ L_0 = R/\xi_0 $ and $ M $ by $\xi$-dependent factors
of order one. Therefore, the galaxy length scale $ L_0 $ and galaxy mass $ M $ emerging from the
Thomas-Fermi equation (\ref{nu}) are a measure of the Jeans' length $ \lambda_J $ 
and the Jeans' mass $ M_J $, respectively.

\medskip

From eqs.(\ref{den2}), (\ref{defalf}) and 
using $ Q(0) = 3 \; \sqrt3 \; \rho(0)/v^3(0) $, eq.(\ref{velgc}) can be rewritten as
\be\label{Qnu0}
\frac{\alpha_0}{\beta_0^{\frac13}} = \frac{\left\{\int_0^{\infty} y^4 \; dy \; 
\Psi[y^2 - \nu(0)]\right\}^{\frac12}}{\left\{\int_0^{\infty} y^2 \; dy \; 
\Psi[y^2 - \nu(0)]\right\}^{\frac56}} =\frac{2.145 \ldots}{\hbar} \; 
\left[\frac{{\rm keV}^4}{Q(0)}\right]^{\frac13} \; \left(\frac{m}{\rm keV}\right)^{\frac43}=
1.0772 \; \left[ \displaystyle \; \frac{M_\odot}{Q(0) \;{\rm pc}^3 \;
\left(\frac{{\rm km}}{{\rm s}}\right)^3} \right]^{\frac13} \; 
\left(\frac{m}{\rm keV}\right)^{\frac43} \; .
\ee
Eq.(\ref{Qnu0}) shows that $ \nu(0) $ is determined by the phase density at the origin
$ Q(0) $.  In the next subsection we solve the Thomas-Fermi eqs.(\ref{nu}) in the whole range of 
the chemical potential at the origin $ \nu(0) $.

\subsection{Physical galaxy properties from the resolution of the Thomas-Fermi equation}

Large positive values of the chemical potential at the origin $ \nu(0) \gg 1 $ correspond to the degenerate 
fermions limit which is the extreme quantum case and oppositely, $ \nu(0) \ll -1 $ gives 
 the diluted limit. The diluted limit is the classical limit and in this case the Thomas-Fermi equations
(\ref{pois})-(\ref{den}) become the equations for a selfgravitating Boltzmann gas.

\medskip

The bounds on the phase space density follows from eqs.(\ref{den2}), (\ref{defalf})
and (\ref{Qab}),
\be                   
0 < \hbar^3 \; \frac{Q(0)}{m^4} \leq  \frac{5 \; \sqrt5}{3 \; \pi^2} = 0.37760\ldots
  \quad {\rm and} \quad 0 < \hbar^\frac32 \; \frac{\sqrt{Q(0)}}{m^2} \leq   0.61449\ldots
\quad {\rm for} \quad  -\infty < \nu(0) \leq \infty
\ee
The largest value for the phase space density corresponds to the degenerate fermions limit
while the smallest values appear in the classical dilute limit. 

\medskip

In the quantum degenerate fermions limit, the halo radius, the velocity dispersion and the galaxy mass 
take their {\it minimum} values:
\bea
&& r_{h \; min} = 24.516 \; {\rm pc} \; \left(\frac{\rm keV}{m}\right)^{\frac43} \; 
\left[\rho(0) \; \frac{{\rm pc}^3 }{M_\odot}\right]^{\frac16} \quad , \cr \cr
&& M_{min} = 1.291 \; 10^5 \; M_\odot \; \left(\frac{\rm keV}{m}\right)^4 \; 
\sqrt{\frac{\rho(0) \; {\rm pc}^3}{M_\odot}}
\quad , \cr \cr
&& v_{min}(0) = 4.768 \; \frac{{\rm km}}{{\rm s}} \; \left(\frac{\rm keV}{m}\right)^{\frac43} \; 
\left[\frac{\rho(0) \; {\rm pc}^3 }{M_\odot}\right]^{\frac13} \; .\label{masam}
\eea
These minimum values are similar to the estimates for degenerate fermions eqs.(\ref{semic})-(\ref{vsemic}), as
it must be.

\medskip

The masses of compact dwarf spheroidal galaxies dominated by DM must be larger than the minimum mass 
$ M_{min} $ eq.(\ref{masam}). The lightest known galaxy of this kind is Willman I (see Table \ref{pgal}).
Imposing $ M_{min} < M_{Willman ~ I} =  3.9 \; 10^5 \; M_\odot $ gives a lower bound for the WDM particle mass:
\be
m > 0.96 \; {\rm keV} \; . 
\ee

Approaching the classical diluted limit yields larger and larger halo radii, galaxy masses
and velocity dispersions. Their maximum values are limited by the initial conditions
provided by the primordial power spectrum which determines the sizes and masses of the galaxies formed.

The phase space density decreases from its maximum value for the compact dwarf galaxies 
corresponding to the degenerate fermions limit till its smallest value for large galaxies
(spirals and ellipticals) corresponding to the classical dilute regime.

Thus, the whole range of values of the chemical potential at the origin 
$ \nu(0) $ , given by the boundary condition, 
from the extreme quantum (degenerate) limit  $ \nu(0) \gg 1 $ to the classical (Boltzmann) dilute
regime $ \nu(0) \ll -1 $ yield all masses, sizes, phase space densities and velocities of galaxies from the 
ultra compact dwarfs till the larger
spirals and elliptical in agreement with the observations (see Table \ref{pgal}).

\medskip

In the degenerate limit the  equilibrium  FD thermal case and the  out of equilibrium case
give identical results, as expected.

\medskip

\begin{figure}[h]
\begin{center}
\begin{turn}{-90}
\psfrag{"3LRQ0sm4.dat"}{Fermions at thermal equilibrium}
\psfrag{"3LRenRQ0sm2.dat"}{Fermions out of thermal equilibrium}
\psfrag{"datoRQ.dat"}{Observed values taking m = 1 keV}
\psfrag{"2datoRQ.dat"}{Observed values taking m = 2 keV}
\psfrag{"5datoRQ.dat"}{Observed values taking m = 5 keV}
\includegraphics[height=13.cm,width=8.cm]{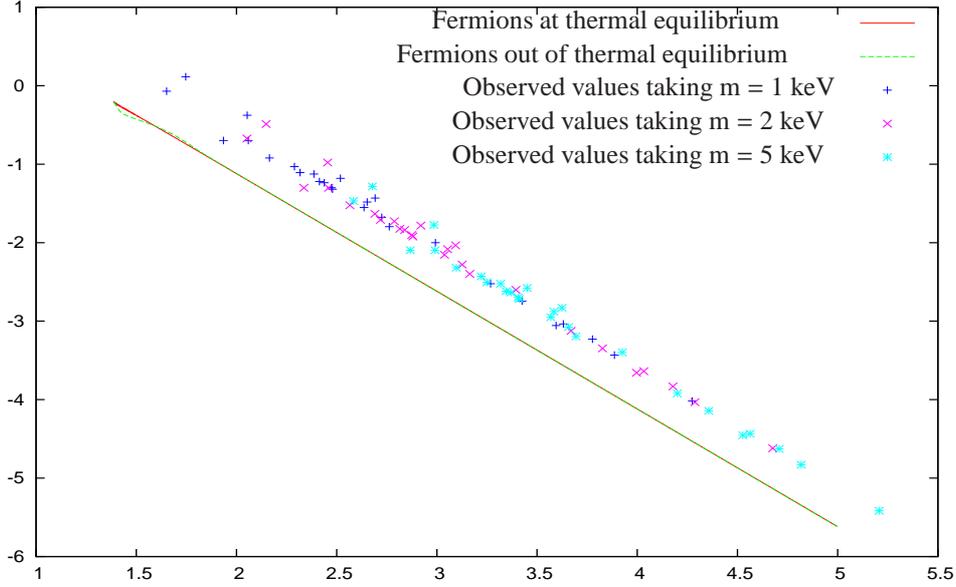}
\end{turn}
\caption{The ordinary logarithm of the square root of the galaxy phase-space density 
$ \log_{10}[\hbar^\frac32 \; \sqrt{Q(0)}/m^2] =  \log_{10}[
3^\frac34  \; \beta_0 /( \pi \; \alpha_0^3 )] $ defined in  eq.(\ref{Qab}) 
as a function of the ordinary logarithm of the product $ \log_{10}\{r_h \; (m/{\rm kev})^\frac43 \; 
[{\rm pc}^3 \; \rho(0)/ M_\odot]^{\frac16} \} = \log_{10}[ R_0 \; \beta_0^{\frac16} \; \xi_h] $ in parsecs 
from the numerical resolution of the Thomas-Fermi eqs.(\ref{nu})
for WDM fermions. The red (solid) curve is for thermal fermions and the green (dashed) curve
corresponds to out of equilibrium sterile neutrinos with the distribution  eq.(\ref{fnue}).
The blue crosses $ + $ are the observed values of $ \hbar^\frac32 \; \sqrt{Q(0)}/m^2 $ from Table \ref{pgal}  
for $ m = 1 $ keV, the red $ X $ are the observed values for $ m = 2 $ keV
and the light blue stars are the observed values for $ m = 5 $ keV. Notice that the observed $ Q_h $ 
from the stars' velocity dispersion are in fact upper bounds for the DM $ Q_h $.}
\label{RQhalo}
\end{center}
\end{figure}

\begin{figure}[h]
\begin{center}
\begin{turn}{-90}
\psfrag{"3Lmrho2.dat"}{Fermions at thermal equilibrium}
\psfrag{"3LRenmrho2.dat"}{Fermions out of thermal equilibrium}
\psfrag{"datoMV.dat"}{Observed values taking m = 1 keV}
\psfrag{"2datoMV.dat"}{Observed values taking m = 2 keV}
\psfrag{"5datoMV.dat"}{Observed values taking m = 5 keV}
\includegraphics[height=13.cm,width=8.cm]{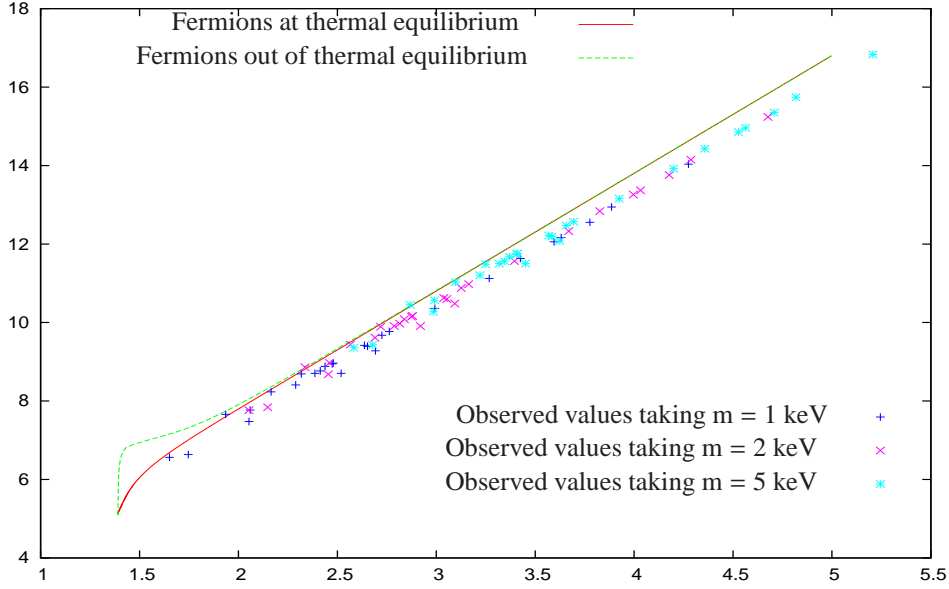}
\end{turn}
\caption{The ordinary logarithm of the galaxy mass 
$ \log_{10}[ (M / M_\odot) \sqrt{M_\odot / [\rho(0) \; {\rm pc}^3]} \; \; \left(m/{\rm keV}\right)^4 ] 
= \log_{10}[1.425 \; 10^5 \; \; \xi_0^2 \;  \left| \nu'(\xi_0) \right|/\sqrt{\beta_0}] $ 
as a function of the ordinary logarithm of the product $ \log_{10}\{r_h \; (m/{\rm kev})^\frac43 \; 
[{\rm pc}^3 \; \rho(0)/ M_\odot]^{\frac16} \} = \log_{10}[R_0 \; \beta_0^{\frac16} \; \xi_h] $ 
in parsecs from the numerical resolution of the Thomas-Fermi eqs.(\ref{nu})
for WDM fermions.  The red (solid) curve is for thermal fermions and the green (dashed) curve
corresponds to out of equilibrium sterile neutrinos with the distribution  eq.(\ref{fnue}). 
The blue crosses $ + $ are the observed values of 
$ \log_{10}[ (M / M_\odot) \sqrt{M_\odot / [\rho(0) \; {\rm pc}^3]} \; \; \left(m/{\rm keV}\right)^4 ] $ 
from Table \ref{pgal} for $ m = 1 $ keV, the red $ X $ are the observed values for $ m = 2 $ keV
and the light blue stars are the observed values for $ m = 5 $ keV.}
\label{Mhalo}
\end{center}
\end{figure}

In fig. \ref{RQhalo} we plot from eq.(\ref{Qab}) the dimensionless quantity 
\be\label{cute1}
\frac{\hbar^\frac32 \; \sqrt{Q(0)}}{m^2} = \frac{3^\frac34}{\pi} \; \frac{\beta_0}{\alpha_0^3} \; .
\ee
In fig. \ref{Mhalo}, we plot the dimensionless product
\be\label{cute2}
\frac{M}{M_\odot} \sqrt{\frac{M_\odot}{\rho(0) \; {\rm pc}^3}} \; \; \left(\frac{m}{{\rm keV}}\right)^4 =
 1.425 \; 10^5 \; \frac{\xi_0^2}{\sqrt{\beta_0}} \; \; \left| \nu'(\xi_0) \right| \; ,
\ee
where $ M $ is the galaxy mass and we used eqs.(\ref{eregc})-(\ref{velgc}). 
In both figures we plot in the abscissa the product 
\be\label{cute3}
r_h  \; \left(\frac{m}{\rm keV}\right)^\frac43 \; \left[\frac{\rm pc^3}{M_\odot} \; \rho(0)\right]^{\frac16}
= R_0 \; \beta_0^{\frac16} \; \xi_h  \quad {\rm in ~ parsecs,}
\ee
where  $ r_h $ is the halo radius. The phase-space density $ Q(0) $ and the galaxy mass $ M $ are obtained 
by solving the Thomas-Fermi eqs.(\ref{nu}) for thermal (FD) fermions and for out of equilibrium sterile neutrinos 
with the distribution eq.(\ref{fnue}).  

Notice that the theoretical curves in the right-hand side of eqs.(\ref{cute1})-(\ref{cute3}) are independent
of the value $ m $ of the DM particle and do not change with $ m $.

\medskip

In fig. \ref{RQhalo} we have superimposed the observed values of $ \sqrt{Q_h}/m^2 $ for $ m = 1 $ keV, $ m = 2 $ keV and
$ m = 5 $ keV (see Table  \ref{pgal}). Notice that the observed values of $ Q_h $ from the stars' velocity dispersion 
are in fact upper bounds for the DM $ Q_h $. This may explain why the theoretical Thomas-Fermi curves in fig. \ref{RQhalo}
appears below the observational data. Notice that the error bars of the observational data are not
reported here but they are at least about $ 10-20 \%$.

\medskip

In fig. \ref{Mhalo} we have superimposed the observed values of  $ (M / M_\odot) \sqrt{M_\odot / 
[\rho(0) \; {\rm pc}^3]} \; \; \left(m/{\rm keV}\right)^4 $  for $ m = 1 $ keV, $ m = 2 $ keV and
$ m = 5 $ keV (see Table  \ref{pgal}). 

\medskip

We see from fig. \ref{RQhalo} that increasing the DM particle value
just pushes down and to the right the observed values $ \hbar^\frac32 \; \sqrt{Q_h}/m^2 $.
In fig. \ref{Mhalo} we see that increasing the DM particle value just pushes
up and to the right the observed values $ (M / M_\odot) \sqrt{M_\odot / [\rho(0) \; {\rm pc}^3]} 
\; \; \left(m/{\rm keV}\right)^4 $.

As noticed above from eqs.(\ref{cute1})-(\ref{cute3}), the theoretical curves are independent 
of the value $ m $ of the DM particle. Hence, for growing $ m \gtrsim $ keV
the left part of the theoretical curves will have no observed galaxy counterpart. Namely, increasing  
$ m \gg $ keV shows an overabundance of small galaxies (small scale structures) without observable counterpart.
This is a further indication that the WDM particle mass is approximately in the range 1 - 2 keV.

\medskip

The galaxy velocity dispersions from eq.(\ref{velgc}) turn to be fully consistent with the
galaxy observations in Table \ref{pgal}.

We see in figs. \ref{RQhalo} and \ref{Mhalo} that fermions at equilibrium and out of  equilibrium
give very similar values for  $ Q(0) $ and $ M $. 
The theoretical values for $ r_h , \; M $ and $ v(0) $ vary very little with the distribution
function $ \Psi $.
This is similar to the WDM linear power spectrum \cite{fluc}, 
where changing the distribution function can be balanced by changing the mass $ m $ within the keV scale.

Notice that the values obtained for the halo radius $ r_h $ in fig. \ref{RQhalo} are much larger than the
quantum lower bound $ r_{min} $ eq.(\ref{rqe}), as expected.

\medskip 

In summary, the theoretical Thomas-Fermi results are fully consistent with all the observations especially
for dwarf compact galaxies as can be seen from Table \ref{pgal}. This result gives an additional support to
the idea put forward in sec. \ref{qup} that galaxies are supported against gravity by the fermionic 
WDM quantum pressure. 

It is highly remarkably that in the context of fermionic WDM the simple static
quantum description provided by Thomas-Fermi is able to reproduce such broad variety of galaxies.

\medskip

To conclude, eqs.(\ref{rgalc})  indicate 
that the galaxy magnitudes  (halo radius, galaxy masses and velocity dispersion) 
obtained from the Thomas-Fermi quantum treatment for fermion masses in the keV scale are
fully consistent with all the observations especially for compact dwarfs (see Table \ref{pgal}). 
Namely, fermionic WDM treated quantum mechanically (as it must be) is able to reproduce
the observed sizes of the DM cores of galaxies.

\section*{acknowledgments}

We are grateful to Peter Biermann for useful discussions in many occasions.
We thank Daniel Boyanovsky for useful remarks.

\appendix

\section{Halo bounds from the Pauli Principle}\label{calK}

We derive here bounds on the phase space density $ Q(r) $ from the Pauli Principle
assuming a simple factorized form for the phase space distribution function
$$
 f({\vec r},{\vec p}) = n({\vec r}) \; \Psi(p/p_0) \; ,
$$
where $ p_0 $ is the momentum characteristic scale of the DM particles.
For fermions at thermal equilibrium $ p_0 $ coincides with the temperature.
Since the DM number density follows integrating $ f({\vec r},{\vec p}) $
over $ \vec p $ as in eq.(\ref{n}), the function $ \Psi(p/p_0) $
must fulfill
$$
1 = \int d^3p \; \frac{\Psi(p/p_0)}{(2 \, \pi \; \hbar)^3} = \frac{p_0^3}{2 \; \pi^2 \; \hbar^3 }
\; \int_0^{\infty} x^2 \; dx \; \Psi(x) 
$$
We can thus write the phase space distribution function as
\be\label{ffact}
 f({\vec r},{\vec p}) = n({\vec r}) \; \frac{2 \; \pi^2 \; \hbar^3 }{p_0^3} \; 
\psi\left(\frac{p}{p_0}\right) \; ,
\ee
where the function 
$$ \psi(x) \equiv \Psi(x) \;  \frac{p_0^3}{2 \; \pi^2 \; \hbar^3} \quad , \quad
x = \frac{p}{p_0} \; ,
$$ 
is normalized by
$$
 \int_0^{\infty} x^2 \; dx \; \psi(x) = 1 \; . 
$$
The Pauli principle imposes the bound eq.(\ref{pauli}) which combined with eq.((\ref{ffact}) yields,
$$
 n({\vec r}) \leq \frac{p_0^3}{\hbar^3 \; \pi^2 \;  \psi(p/p_0)} \; .
$$
Since the momentum distribution $ \psi(p/p_0) $ is normally a monotonically
decreasing function of $ p $, the most stringent bound is obtained setting $ p = 0 $:
\be
 n({\vec r}) \leq \frac{p_0^3}{\hbar^3 \; \pi^2 \;  \psi(0)} = 
\frac{\sqrt{27} \; m^3 \; \sigma^3}{\hbar^3 \; \pi^2 \;  \psi(0)} \; .
\ee
Therefore, the phase space density eq.(\ref{cotQ}) is bounded by
\be
Q({\vec r}) \leq K \; \frac{m^4}{\hbar^3}  \; ,
\ee
where the dimensionless quantity $ K $ is given by
\be \label{defK}
K \equiv \frac{\sqrt{27}}{\pi^2 \;  \psi(0)}
\ee
The value of $ K $ depends on the nature of the momentum distribution.

\bigskip

For DM particles in thermal equilibrium $ p_0 $ is equal to the temperature $ T $ and 
\be\label{fd}
\Psi(x) = \frac1{\displaystyle e^x + 1} \; .
\ee
For sterile neutrinos in the $\nu$-MSSM model \cite{numssm} which decouple out of equilibrium.
Their  freezed-out distribution function $ \Psi(x) $ is given by \cite{dest}
\be \label{fnue}
\Psi(x) = 2 \; \tau \; \sqrt{\frac{\pi}{x}} \sum_{n=1}^{\infty} \frac{e^{-n \, x}}{n^{\frac52}} \; .
\ee
where $ \tau \simeq 0.03 $ is a coupling constant. This formula is valid for all 
$ \sqrt{x} > \tau $ and we take $ \Psi(x) = 1 $ for $ \sqrt{x} < \tau $. 

\medskip

For sterile neutrinos in the Dodelson-Widrow model \cite{DW} we have (approximately)
the  freezed-out distribution function
\be\label{dw}
\Psi(x) = \frac{f_0}{m} \; \frac1{\displaystyle e^x + 1} \quad {\rm where}
\quad f_0 \simeq 0.043  \; {\rm keV} \; .
\ee

\medskip

We display  $ K $ in Table \ref{valK}  for thermal fermions, out of thermal
equilibrium fermions in the DW and $\nu$-MSM models and for the Maxwell-Boltzmann
distribution. 

\medskip

Notice that the quantum bound is more restrictive ($ K $ turns to be smaller)
in the out of thermal equilibrium $\nu$-MSM model than in the other cases.
This is due to the fact that generically, out of equilibrium distributions
as the  $\nu$-MSM
have more particles with low momentum \cite{ddv1,ddv2}. $ \psi(0) $ is therefore larger 
than at thermal equilibrium and from eq.(\ref{defK}) $ K $ is smaller.

\medskip

The equilibrium Fermi-Dirac distribution produces the less restrictive
bound in Table  \ref{valK} because, somehow, it already knows about
the exclusion principle.

\end{document}